\documentstyle[aps,epsfig,prd]{revtex} 

\draft

\begin{document}

%@twocolumnfalse\endcsname 

\title{Cosmic Neutrinos and New Physics beyond the Electroweak Scale}
\author{Craig~Tyler, Angela V. Olinto}
\address{Department of Astronomy \& Astrophysics,
Enrico Fermi Institute, The University of Chicago, Chicago, 
IL~~60637-1433}

\author{G\"unter~Sigl} 
\address{DARC, UMR--8629 CNRS, Observatoire de Paris-Meudon, F--92195
Meudon C\'edex, France}

\maketitle

\begin{abstract}
New physics beyond the electroweak scale may increase
weak interaction cross sections beyond the Standard Model
predictions.  Such cross sections can be
expected within theories that solve the hierarchy problem
of known interactions with a unification scale in the TeV
range. We derive constraints on these cross sections from
the flux of neutrinos expected from cosmic ray interactions
with the microwave background and the non-observation
of horizontal air showers. We also discuss how this limit can
be improved by upcoming cosmic ray and neutrino experiments,
and how the energy dependence of the new interactions can be
probed by these experiments.
\end{abstract}


%\vskip2.2pc]

\section{Introduction}

It has been suggested that the neutrino-nucleon
cross section, $\sigma_{\nu N}$, can be enhanced by new
physics beyond the electroweak scale in the center of
mass (CM) frame, or above about a PeV in the nucleon rest frame.
At ultra-high energies (UHE),  neutrinos can in principle acquire $\nu N$
cross sections approaching hadronic levels.  It
may therefore be possible for UHE neutrinos to initiate air
showers, which would offer a very direct means of probing
these new UHE interactions.  The results of this kind of
experiment have important implications for both neutrino
astronomy and high energy physics.

For the lowest partial wave contribution to the cross section
of a point-like particle, this new physics would
violate unitarity~\cite{bhg}. However, two major possibilities
have been discussed in the literature for which unitarity
bounds need not be violated. In the first, a broken SU(3)
gauge symmetry dual to the unbroken SU(3) color gauge group
of strong interaction is introduced as the ``generation symmetry'' such
that the three generations of leptons and quarks represent the quantum
numbers of this generation symmetry. In this scheme, neutrinos can have
close to strong interaction cross sections with quarks.
In addition, neutrinos can
interact coherently with all partons in the nucleon, resulting in
an effective cross section comparable to the geometrical
nucleon cross section.  This model lends itself to experimental
verification through shower development altitude statistics, as
described by its authors~\cite{bhfpt}.  The present paper will not
affect the plausibility of this possibility, which largely awaits
more UHE cosmic ray (UHECR) events to compare against.

The second possibility consists of a large increase
in the number of degrees of freedom above the electroweak 
scale~\cite{kovesi-domokos}. A specific implementation
of this idea is given in theories with $n$ additional large
compact dimensions and a quantum gravity scale $M_{4+n}\sim\,$TeV
that has recently received much attention in the
literature~\cite{tev-qg} because it provides an alternative
solution (i.e., without supersymmetry) to the hierarchy problem
in grand unifications of gauge interactions.
For parameters consistent with measured cross sections at
electroweak energies and below, at $\simeq10^{20}\,$eV, the
$\nu N$ cross section can approach $10^{-4}-10^{-2}$ times
a typical strong cross section~\cite{jain}. Therefore,
in these models the mean free path for $\nu N$ interactions
at these energies is on the order of kilometers to hundreds
of kilometers in air. Experimentally, this would be
reflected by a specific energy dependence
of the typical column depth at which air showers initiate.
As we will show, this leads to a useful signature, deriving from zenith
angle variations with shower energy.  Comparison with
observations would either support these scenarios or constrain
them in a way complementary to studies in terrestrial
accelerators~\cite{coll-extra-dim} or other laboratory
experiments that have recently appeared in the literature.

Both scenarios for increasing $\nu N$ cross sections raise the question
of how far UHE neutrinos can travel in space.  In the first case,
due to the {\it coherent} nuclear interactions, the cross section
approaches the nucleon-nucleon cross section, and so intergalactic
matter becomes
the greatest threat to the neutrino traveling long distances
unimpeded.  However, an order of magnitude calculation shows
that the mean free path for such a neutrino is $\sim10^2$~Gpc,
if we simply assume the matter density of the Universe is
its critical density.  Within a galaxy with similar interstellar
density to the Milky Way, the mean free path is $\sim\,$Mpc;
the UHE neutrino is not attenuated before reaching Earth's atmosphere.
For the extra dimension model, the cross sections are lower, so
interaction with matter is not a problem.  With Cosmic Microwave
Background (CMB) photons, the CM energy is
too low to give rise to a significant interaction
probability. If neutrinos are massive,
then the relic neutrino background is more of a threat.  But
assuming the neutrino mass is below 92 eV, as conservatively
required to prevent over-closing the Universe for a stable
species~\cite{kolb}, the mean free path is $\sim10^4$~Gpc
at a CM energy of $\sim100\,$GeV, corresponding to a
$\sim10^{20}\,$eV neutrino.
The cosmological horizon is only $\sim15$~Gpc, so for both
scenarios the neutrinos can reach Earth unhindered from
arbitrarily great distances.

A major motivation for attempts to augment the $\nu N$ cross section
at ultra-high energies derives from the paradox of UHECRs.
Above $\sim10^{20}$~eV, protons and neutrons see CMB photons
doppler shifted to high energy gamma rays, and encounters
generate pions, thereby reducing the resulting nucleon's
energy by at least the pion rest mass.  One therefore expects
the Greisen-Zatespin-Kuzmin (GZK) cutoff~\cite{gzk}, wherein cosmic
rays above $E_{\rm GZK}\simeq7\times10^{19}$~eV would lose
their energy within $\simeq50\,$Mpc.
However, since we observe cosmic rays at super-GZK energies,
it has been proposed that these UHECRs might really be neutrinos
rather than nucleons (for example, see~\cite{bhfpt}).  While
this is a strong motivation to explore new UHE neutrino
interactions, we will not need to assume that UHECRs are neutrinos.

The paper is outlined as follows: In Section~II we derive
a constraint on the neutrino-nucleon cross section
resulting from the non-observation of horizontal
air showers induced by neutrinos produced by interactions
of the highest energy cosmic rays with the cosmic microwave
background. In Section~III we discuss potential improvements
of this constraint including other hypothesized neutrino
sources and next generation experiments. In Section~IV
we narrow the discussion to the case of a $\nu N$ cross
section scaling linearly with CM energy, as an example. 
A zenith angle profile is presented as a signature of
this scaling, and a means of constraining it.
We conclude in Section~V.

\section{A Bound from the ``Cosmogenic'' Neutrino Flux}

The easiest way to distinguish extensive air showers (EAS)
caused by neutrinos from those caused by hadrons is
to look for nearly horizontal electromagnetic showers.
As pointed out in Ref.~\cite{auger-proc}, the Earth atmosphere is
about 36 times thicker taken horizontally than vertically.
A horizontally incident UHE hadron or nucleus has a vanishingly
small chance of descending to sufficiently low altitudes to
induce air showers detectable by ground
instruments.   As a consequence, a horizontal shower is a
very strong indication of a neutrino primary.

The two major techniques for observing extensive air showers
produced by UHECRs or neutrinos are the
detection of shower particles by an array
of ground detectors, and the detection of the nitrogen
fluorescence light produced by the shower. The largest
operating experiment utilizing the first technique
is the AGASA which covers about $100\,{\rm km}^2$
in area~\cite{agasa}. The second technique was pioneered
by the Fly's Eye instrument~\cite{fe} and produced
a total exposure similar to the AGASA instrument.  For
cosmic neutrino detection, the fluorescence technique
may provide somewhat more detailed information, because it
can readily measure the position of the shower maximum and
possibly the first interaction, which are important in
distinguishing neutrino primaries.  Also, fluorescence
detectors are equipped to see horizontal showers
that do not intersect the ground near the detector.

The main backgrounds consist of muon bremsstrahlung and
tau lepton decays, as these could occur very deep in
the atmosphere.  The flux of these particles is several orders
of magnitude below the CRs and predicted neutrinos of the
same energy, but if UHE neutrinos interact with matter only
through Standard Model channels, then the rate of
neutrino-induced triggers will still be dwarfed by these
background events.  The backgrounds for high zenith angle
events differ somewhat between fluorescence detectors and
ground arrays~\cite{baltrusaitis,auger-muon}, but both detector
classes have remedies for the high background.  For example,
one interesting benefit with air fluorescence is the possibility
of seeing a separate primary CR shower associated with the event,
thus identifying it as a muon- or tau-induced
secondary~\cite{yoshida1}.  Ground arrays would not see this sign,
but they should be able to discriminate between neutrinos and
secondaries by the shape of the shower front~\cite{auger-gap}.

The strategy here is to use observations (or non-observation)
of horizontal showers to place limits on the UHE neutrino
flux incident on the Earth.  Comparing that against a reliably
predicted flux permits us to bound the cross section.

\begin{figure}
\centering\leavevmode
\epsfig{file=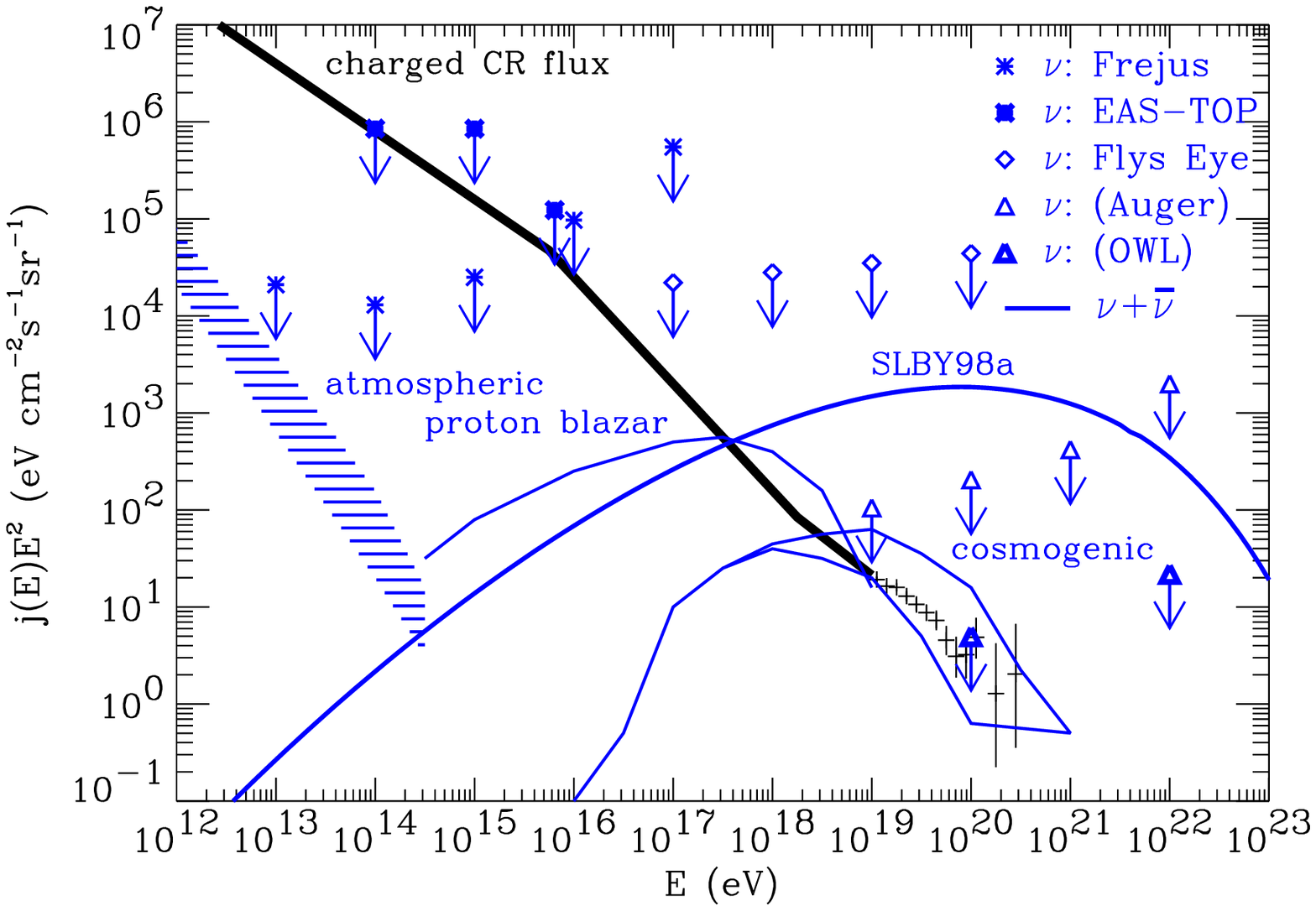,width=6.0in,angle=0}
\vskip1cm
\caption[...]{Predictions for the summed differential fluxes of all
neutrino flavors (solid lines) from the atmospheric
background for different zenith angles~\cite{lipari} (hatched
region marked ``atmospheric''), from UHECR interactions with
the CMB~\cite{pj} (the two curves labeled ``cosmogenic''
indicating a typical uncertainty range), from proton blazars that are
photon optically thick to nucleons but contribute to the diffuse
$\gamma-$ray flux~\cite{protheroe2}
(``proton blazar''), and for a model where UHECRs are produced
by decay of particles close to the Grand Unification Scale
(``SLBY98'', see Ref.~\cite{slby} for details).
1 sigma error bars are shown on the combined data from the
Haverah Park~\cite{haverah},
the Fly's Eye~\cite{fe}, and the AGASA~\cite{agasa} experiments
above $10^{19}\,$eV. Also shown are piecewise power law fits to the
observed
charged CR flux (thick solid line).
Points with arrows represent approximate upper limits on the
diffuse neutrino flux from the Frejus~\cite{frejus}, the
EAS-TOP~\cite{eastop2}, and the Fly's
Eye~\cite{baltrusaitis} experiments, as indicated. The projected
sensitivity for the Pierre Auger project uses the acceptance
estimated in Ref.~\cite{auger-neut}, and the one for the OWL concept
study is based on Ref.~\cite{owl}, both assuming observations
over a few years period.  Note that this plot supposes that
$\bar y=1$; for $\bar y<1$, data points correspond to an x-axis
of observed shower energy, and the predicted flux curves
correspond to an x-axis of incident primary energy.  If the x-axis
is to be interpreted as observed shower energy for $\bar y<1$,
then the predicted flux curves should be shifted to the left
by a factor of $\bar y$ and multiplied by $\bar y$.
\label{flux}}
\end{figure}

Fig.~\ref{flux} summarizes the high energy neutrino fluxes
expected from various sources, where the shown flavor summed
fluxes consist mostly of muon and electron neutrinos in a ratio
of about 2:1. Shown are the atmospheric
neutrino flux~\cite{lipari}, a typical flux range expected for
``cosmogenic'' neutrinos created as secondaries from the decay
of charged pions produced by collisions between UHE nucleons and
CMB photons~\cite{pj}, a typical prediction for the
diffuse flux from photon optically thick proton
blazars~\cite{protheroe2} that were normalized to recent
estimates of the blazar contribution to the diffuse $\gamma-$ray
background~\cite{muk-chiang}, and a typical
prediction by a scenario where UHECRs
are produced by decay of particles close to the
Grand Unification Scale~\cite{slby} (for a review see
also Ref.~\cite{bs}).

Apart from the atmospheric neutrino flux, only the cosmogenic
neutrinos are nearly guaranteed to exist due to the known existence
of UHECRs, requiring only that these contain nucleons and be
primarily extragalactic in origin.  This is an assumption,
since Galactic sources can not be wholly ruled out~\cite{olinto}.
However, if the charge of the primary UHECR satisfies
$Z\lesssim{\cal O}(1)$ (which will be tested by forthcoming
experiments), an extragalactic origin is favored on the purely
empirical grounds of the observed nearly isotropic angular
distribution which rules out Galactic disk
sources.  The reaction generating these cosmogenic neutrinos is
well known, depending only on the existence of UHE nucleon sources more
distant than $\lambda_{\rm GZK}\simeq$~8 Mpc and on the relic microwave
photons known to permeate the Universe.
The two cosmogenic lines in Fig.~\ref{flux} were
computed for a scenario where radiogalaxies explain UHECRs
between $\simeq3\times10^{18}\,$eV (i.e. above the ``ankle'')
and $\simeq10^{20}\,$eV~\cite{rb},
with source spectra cutoffs at $3\times10^{20}$~eV (lower curve) and
$3\times10^{21}$~eV (upper curve)~\cite{pj}. Since events
were observed above $3\times10^{20}\,$eV, this may be used as a
conservative flux estimate (see also Sect.~III below).

The non-observation of horizontal showers by the neutrino fluxes indicated
in Fig.~\ref{flux} can now be translated into an upper limit on
the total neutrino-nucleon cross section. The total charged
current Standard Model $\nu N$ cross section can be approximately
represented by~\cite{gqrs}
\begin{equation}
\sigma^{SM}_{\nu N}(E)\simeq2.36\times10^{-32}(E/10^{19}
  \,{\rm eV})^{0.363}\,{\rm cm}^2\quad(10^{16}\,{\rm eV}\lesssim
  E\lesssim10^{21}\,{\rm eV})\,.\label{cross}
\end{equation}

In any narrow energy range and as long as the neutrino mean free
path is larger than the linear detector size, the neutrino
detection rate scales as~\cite{auger-proc}
\begin{equation}
{\rm Event~Rate}\propto\,A\,\phi_{\nu}\,\sigma_{\nu N}\,,
\label{event_rate}
\end{equation}
where $A$ is the detector acceptance (in units of volume times
solid angle) in that energy band, and
$\phi_{\nu}$ is the neutrino flux incident on the Earth.
In our case, the non-observation of neutrino-induced
(horizontal) air showers limits the event rate, and thereby sets
the experimental upper bounds given in Fig.~\ref{flux}.
On the other hand, the gap between the best current experimental flux
bound and the predicted cosmogenic flux specifies a bounding
cross section in the presence of new interactions.  Representing
the experimental bounds in Fig.~\ref{flux} with an approximate
expression, and letting $\bar y$ be the average fraction of the neutrino's
energy deposited into the shower, the cross section limit becomes
\begin{equation}
\sigma_{\nu N}(E)\lesssim10^{-28}\left(\frac{10^{-18}\,
{\rm cm^{-2}s^{-1}sr^{-1}}}{\phi_c(E)}\right)
\left(\frac{10^{19}\,{\rm eV}}{\bar y E}\right)^{1/2}\,
{\rm cm}^2\,,\label{crosslimy}
\end{equation}
where $\phi_c(E)$ is the cosmogenic neutrino flux, and the exponent
1/2 reflects the approximate energy dependence of the Fly's Eye
upper limit on the rate of horizontal air showers.
For electron neutrinos and atmospheric detectors,
charged current (CC) interactions correspond to $\bar y=1$
since all of the neutrino energy goes into ``visible'' energy,
either in the form of an electron or of hadronic debris.
In this case, using the conservative
lower estimate of the cosmogenic flux in Fig.~\ref{flux}
yields
\begin{eqnarray}
  \sigma_{\nu N}^{CC}(10^{17}\,{\rm eV})&\lesssim&1\times10^{-29}
  \,{\rm cm}^2\nonumber\\
  \sigma_{\nu N}^{CC}(10^{18}\,{\rm eV})&\lesssim&8\times10^{-30}
  \,{\rm cm}^2\nonumber\\
  \sigma_{\nu N}^{CC}(10^{19}\,{\rm eV})&\lesssim&5\times10^{-29}
  \,{\rm cm}^2\,,\label{crosslim1}
\end{eqnarray}
and a severely degraded limit outside this energy range.
The bounds Eq.~(\ref{crosslim1}) probe CM energies
\begin{equation}
  \sqrt{s}\simeq1.4\times10^5\left(\frac{E}{10^{19}\,{\rm eV}}
  \right)^{1/2}
  \,{\rm GeV}
\end{equation}
that are about 3 orders of magnitude beyond the
electroweak scale.

In the case of neutral currents, one needs a model for
computing $\bar y$. As an example, if $\bar y=0.1$ at these energies,
then the best limit becomes $\sigma_{\nu N}(10^{18}\,{\rm eV})
\lesssim2\times10^{-29}\,{\rm cm}^2$.  So as long as $\bar y$
doesn't depart too far from $\sim0.1$,
as indicated in~\cite{agmprs},
neutral currents won't
change the bounds dramatically.  In Section IV, we
discuss the neutral current case with respect to extra
dimension models.
 
Note that this bound does not challenge the model of Ref.~\cite{bhfpt}.
That option relies on neutrinos acquiring larger, hadron-scale
cross sections.  As such, shower development will occur vertically,
not horizontally, and our argument does not apply.  We therefore
specify a sub-hadron-size cross section above which our
bound is inapplicable because horizontal air showers could
not develop in column depths $\gtrsim3000\,{\rm g}/{\rm cm}^2$
as used by the Fly's Eye bound~\cite{baltrusaitis}:
\begin{equation}
\sigma_{\nu N}(E\gtrsim10^{19}\,{\rm eV})\gtrsim10^{-27}
\,{\rm cm}^2\,.
\label{crosslim2}
\end{equation}

The bound derived here is independent of the type of
interaction enhancing physics, stemming directly from
attempts to observe deeply penetrating air showers.
To summarize, Eq.~(\ref{crosslimy}) applies
to the cosmogenic neutrino flux as long as $\sigma_{\nu N}$ 
does not reach hadronic levels.

We note that our allowed range, Eqs.~(\ref{crosslim1})
and~(\ref{crosslim2}), is complementary to and  consistent with
the interpretation of UHECRs as neutrino primaries, as
considered in Ref.~\cite{jain}. In this case, 
$\sigma_{\nu N}(E)\gtrsim2\times10^{-26}\,{\rm cm}^2$,
which is in the larger cross section regime, Eq.~(\ref{crosslim2}).
This is because for $\sigma_{\nu N}(E)\lesssim10^{-28}\,{\rm cm}^2$,
the first interaction point would have a flat distribution
in column depth up to large zenith angles $\theta\lesssim80^\circ$,
whereas observed events appear to peak at
column depths $\sim450\,{\rm g}/{\rm cm}^2$~\cite{baltrusaitis}.
For an individual event, a $\sim0.2$\% probability exists
for interacting by $450\,{\rm g}/{\rm cm}^2$ with cross sections
as small as $10^{-29}\,{\rm cm}^2$.

We also note in this context that Goldberg and Weiler~\cite{gw}
have related the $\nu N$ cross section at UHEs to the lower
energy $\nu N$ elastic amplitude in a model independent way,
only assuming $3+1$ dimensional field theory. Consistency
with accelerator data then requires
\begin{equation}
\sigma(E)\lesssim3\times10^{-24}\left(\frac{E}{10^{19}\,{\rm eV}}
\right)\,{\rm cm}^2\,,\label{gwbound}
\end{equation}
otherwise deviations of neutrino cross sections from the
Standard Model should become visible at the electroweak scale~\cite{gw}.

\section{Future Prospects for Improvement}

EAS detection will be pursued by several
experiments under construction or in the proposal stage.
As an upscaled version of the original Fly's Eye Cosmic Ray experiment,
the High Resolution Fly's Eye detector~\cite{hires} has recently
begun operations and has preliminary results of several super-GZK
events~\cite{icrc26}, but no official limit on horizontal showers thus
far. The effective aperture of this instrument
is $\simeq350 (1000)\,{\rm km}^2\,{\rm sr}$ at $10 (100)\,$EeV,
on average about 6 times the Fly's Eye aperture, with a
threshold around $10^{17}\,$eV.
This takes into account a duty cycle of about 10\% which is
typical for the fluorescence technique. Another project utilizing
this technique is the proposed Japanese Telescope
Array~\cite{tel_array}. If approved, its effective aperture will
be about 10 times that of Fly's Eye above $10^{17}\,$eV.
The largest project presently under construction is the  Pierre Auger
Giant Array Observatory~\cite{auger} planned for two sites, one in
Argentina and another in the USA for maximal sky coverage. Each site
will have  a $3000\,{\rm km}^2$ ground array. The southern
site will have about 1600 particle detectors (separated by 1.5 km
each) overlooked by four fluorescence
detectors. The ground arrays will have a duty cycle of nearly 100\%,
leading to an effective aperture about 30 times as large as the AGASA
array. The corresponding cosmic ray event rate above
$10^{20}\,$eV will be 50--100 events per year. About 10\% 
of the events will be detected by both the ground array
and the fluorescence component and can be used for cross
calibration and detailed EAS studies. The energy threshold will
be around $10^{18}\,$eV, with full sensitivity above $10^{19}\,$eV.

NASA recently initiated a concept study for detecting EAS
from space~\cite{owl} by observing their fluorescence light
from an Orbiting Wide-angle Light-collector (OWL). This would
provide an increase by another factor of $\sim50$ in aperture
compared to the Pierre Auger Project, corresponding to a
cosmic ray event rate of up to a few thousand events per year above
$10^{20}\,$eV. Similar concepts being proposed are the
AirWatch~\cite{linsley} and Maximum-energy Air-Shower Satellite
(MASS)~\cite{mass} missions. The energy
threshold of such instruments would be between $10^{19}$ and
$10^{20}\,$eV. This technique would be especially suitable for detection
of almost horizontal air showers that could be caused by UHE neutrinos.
As can be seen from Fig.~\ref{flux}, with an experiment such
as OWL, the upper limit on the cross section Eq.~(\ref{crosslim1})
could improve by about 4 orders of magnitude.
In addition, for such showers the fluorescence technique
could be supplemented by techniques such as radar echo
detection~\cite{gorham}.

The study of horizontal air showers nicely complements the more traditional
technique for neutrino detection in underground detectors. 
In these experiments, muons created in charged current reactions of
neutrinos with nucleons either in water or in ice
are detected via the optical Cherenkov light emitted. Due to
the increased column depth (see also Fig.~\ref{reno}),
this technique would be mostly sensitive
to cross sections smaller by about a factor of 100, in a range
$\sim10^{-31}-10^{-29}\,{\rm cm}^2$. For an event rate
comparable to the one discussed above for ground based instruments, the
effective area of underground detectors needs to be comparable and  
thus significantly larger than $1\,{\rm km}^2$, with
at least some sensitivity to downgoing events.
The largest pilot experiments in ice (AMANDA~\cite{amanda}) 
and in water (Lake Baikal~\cite{baikal} and the now defunct
DUMAND~\cite{dumand}) are roughly
$0.1\,$km in size. Next generation deep sea projects
such as the ANTARES~\cite{antares}
and NESTOR~\cite{nestor} projects in the Mediterranean,
and ICECUBE~\cite{icecube} in Antarctica will significantly improve the
volume studied but may not reach the specific requirements for the tests
discussed here. Also under consideration are techniques
for detecting neutrino induced showers in ice or water
acoustically or by their radio emission~\cite{price}.
Radio pulse detection from the
electromagnetic showers created by neutrino
interactions in ice~\cite{radio-technique}
could possibly be scaled up to an effective area of
$10^4\,{\rm km}^2$ and may be the most promising option
for probing the $\nu N$ cross section with water or ice detectors.
A prototype is represented by the
Radio Ice Cherenkov Experiment (RICE) experiment at the
South Pole~\cite{rice}.

The bounds we derived here can be even more constraining if the 
UHE neutrino fluxes are higher than the conservative estimates we used.
For instance, the cosmogenic flux was computed~\cite{pj}
by assuming a cosmic ray spectral dependence of
$N(E)\propto E^{-\gamma}$ with $\gamma=3$, consistent with
the overall cosmic ray spectrum above $\sim10^{15}$~eV.
However, at $\gtrsim10^{19.5}$~eV, the spectrum appears to
flatten down to $1\lesssim\gamma\lesssim2$, probably
suggestive of a new cosmic ray component at
these energies. Once more UHECR events are observed, the trend hinted by
the AGASA data may be confirmed, giving the source of the UHECRs that
generate the cosmogenic neutrinos a much harder spectral index than we
assumed. The UHE cosmogenic neutrino flux computed for this
flattened spectrum would yield a higher flux  and a stronger bound will
result.  For example, assuming a rather strong source evolution and
an $E^{-1.5}$ injection spectrum~\cite{yoshida1} would increase the
cosmogenic flux up to 100 fold, and consequently yield a 100
fold improvement of the constraint Eq.~(\ref{crosslim1}).
See also Ref.~\cite{sdss} for an estimate more optimistic by a
factor 10-20 than our conservative one.

Concerning sources of primary UHE neutrinos, current
estimates on the flux from the proton blazar model~\cite{mpr},
wherein AGN accelerate protons which interact with the
local thick photon field, producing pions and therefore
neutrinos, could potentially improve the bound by a factor
of $\sim50$ at $10^{17}$~eV.  The topological defect model
shown in Fig.~\ref{flux}~\cite{slby} could improve the bound
by a factor of $\sim75$ at $10^{19}$~eV. These improvements
would be realized if either model were to become as strongly
motivated as the cosmogenic flux.  Furthermore, any mechanism
predicting fluxes approaching the experimental flux
limits would exclude a new contribution to the $\nu N$ cross
section beyond the Standard Model at corresponding energies.
Particularly high UHE neutrino fluxes are predicted in scenarios where
the highest energy cosmic rays are produced as secondaries
from Z bosons resonantly produced in interactions of
these neutrino primaries with the relic neutrino
background~\cite{weiler,roulet,yoshida,yoshida1,slby}.
In general, any confirmed increase in
flux over cosmogenic as currently estimated~\cite{pj} would
improve the bound presented here.

We also consider Markarian 421 as an example of how a point
source can impact the bound.  Its calculated neutrino flux is 
$\sim50~{\rm eV~cm^{-2}~s^{-1}}$ at its peak energy of 
$\sim10^{18}$~eV~\cite{halzen}.  This is about an order of
magnitude shy of the cosmogenic flux at the same energy,
and therefore it will not serve to improve the bound.
However, should some new evidence arise against the
existence of a cosmogenic flux, point source estimates
like this may become the fallback means to constrain $\sigma_{\nu N}$.

\section{Energy Dependence Signatures and Extra Dimension Models}

In extra dimension scenarios, the virtual exchange of bulk gravitons
(Kaluza-Klein modes) leads to extra contributions to any two-particle
cross section. For CM energy such that $s\lesssim M_{4+n}^2$, where
possible stringy effects are under control~\cite{dm}, these cross
sections can be well approximated perturbatively. In contrast, for
$s\gtrsim M_{4+n}^2$ model dependent string excitations can become
important and in general one has to rely on extrapolations
which can be guided only by general principles such as
unitarity~\cite{jain}. Naively, the exchange of spin 2 bulk
gravitons would predict an $s^2$ dependence of the cross
section~\cite{ns,jain}. However, more conservative arguments
consistent with unitarity~\cite{ns,jain} suggest a linear growth in
$s$, and in what follows we will assume the following cross section
parameterization in terms of $M_{4+n}$:
\begin{equation}
  \sigma_g\simeq\frac{4\pi s}{M^4_{4+n}}\simeq
  10^{-28}\left(\frac{{\rm TeV}}{M_{4+n}}\right)^4
  \left(\frac{E}{10^{19}\,{\rm eV}}\right)\,{\rm cm}^2\,,
  \label{sigma_graviton}
\end{equation}
where the last expression applies to a neutrino
of energy $E$ hitting a nucleon at rest. It should be
stressed that this cross section is only one example
among a few proposed in the literature.
We further note that within a string theory context,
Eq.~(\ref{sigma_graviton}) can only be a good
approximation for $s\lesssim M_s^2$.  Above the string
scale, there is currently no agreement in the literature
about the behavior of $\sigma_g$ with energy.
Amplitudes for single states may be exponentially
suppressed $\propto\exp[-s/M_s^2]$ above the string
scale $M_s\sim M_{4+n}$~\cite{dm}, which can
be interpreted as a result of the finite spatial extension
of the string states, in which case the number of states may
grow exponentially~\cite{kovesi-domokos}.  This is currently
unclear, and which effect dominates the total cross
section may be model dependent.  For example, domination
by the amplitude suppression would result in
$\sigma_g\lesssim4\pi/M_s^2\lesssim10^{-32}\,{\rm cm}^2$,
conservatively assuming $M_s\gtrsim100\,$GeV. Thus,
an experimental detection of the signatures discussed
in this section could lead to constraints on some
string-inspired models of extra dimensions.

\begin{figure}
\centering\leavevmode
\epsfig{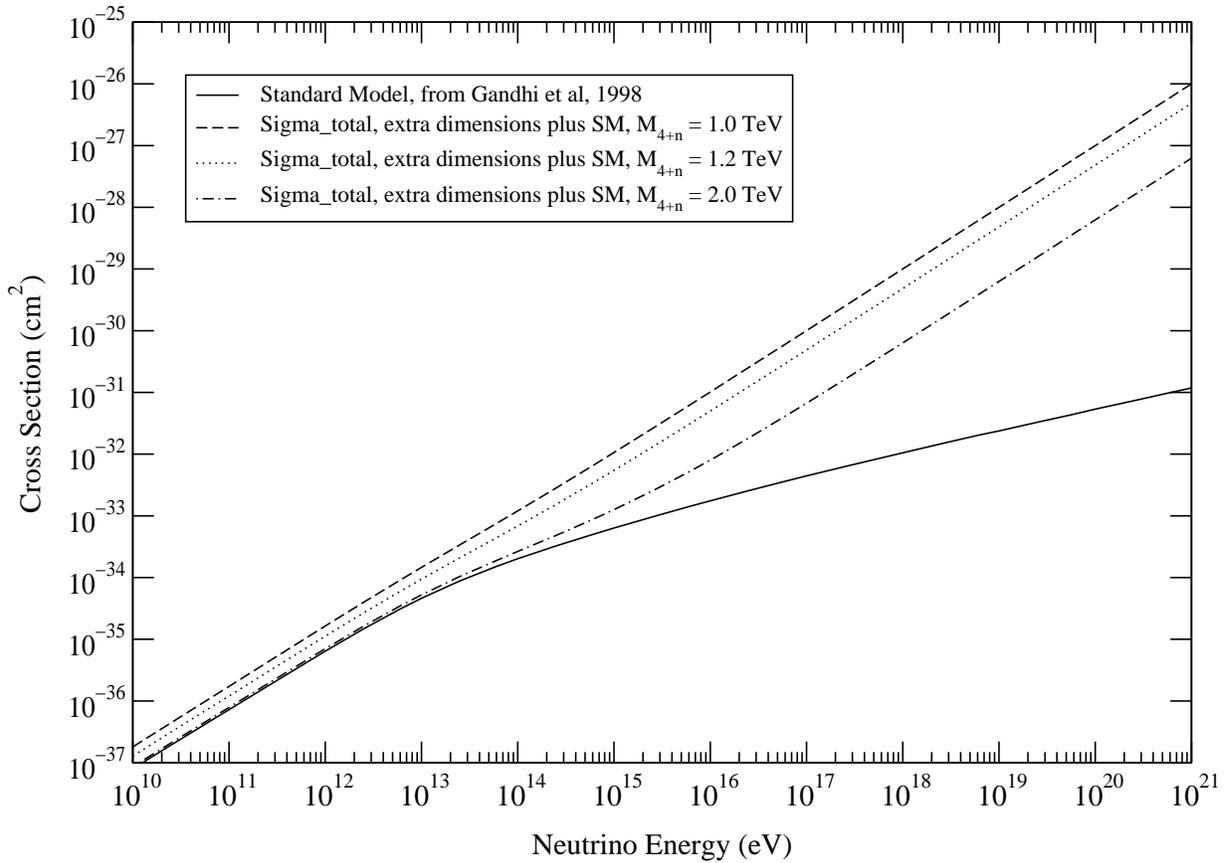}
\vskip1cm
\caption[...]{Cross section for $\nu N$ interaction, assuming
neutrinos incident on nucleons at rest. The solid curve is for the 
Standard Model, charged current~\cite{gqrs}, as appropriate for
electron neutrino detection (see text).  Remaining curves
give Standard Model plus the contribution from extra dimensions
at various scales in the parameterization of
Eq.~(\ref{sigma_graviton}):  dashed line for
$M_{4+n}=1.0$~TeV, dotted line for $M_{4+n}=1.2$~TeV, dot-dashed line
for $M_{4+n}=2.0$~TeV.
\label{sigma}}
\end{figure}

Fig.~\ref{sigma} shows the neutrino-nucleon cross section based on
the Standard Model~\cite{gqrs}, and three curves for enhanced cross
sections, given by the sum
\begin{equation}
  \sigma_{tot}=\sigma_{SM}+\sigma_g\,.
  \label{sigma_sum}
\end{equation}
These three curves are given for different values of the
scale $M_{4+n}$. An increase in this mass scale brings
the total cross section closer to that of the Standard
Model, Eq.~(\ref{cross}).  For electron neutrinos,
the Standard Model contribution is well approximated
(within about 10\%) by the charged current cross section
Eq.~(\ref{cross}), whereas the other flavors are harder
to observe~\cite{auger-gap}.  Of course, experimentally speaking, 
what counts for observing an interaction is
the total cross section weighted by the average energy
fraction transferred to the shower. 

\begin{figure}
\centering\leavevmode
\epsfig{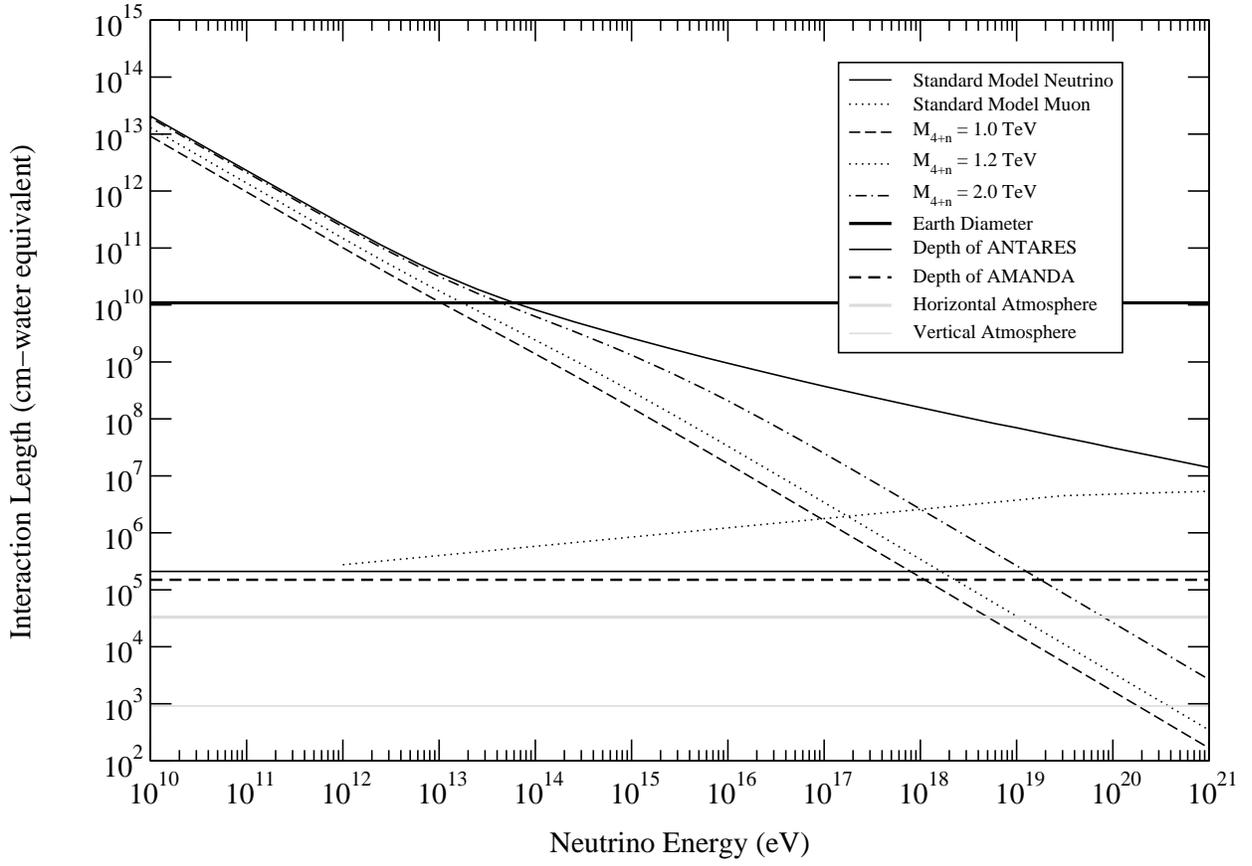}
\vskip1cm
\caption[...]{Interaction length for $\nu N$ interaction in water
or ice, assuming neutrinos incident on nucleons at rest, similar to
plots in Ref.~\cite{gqrs}.  Down-sloping diagonal lines correspond to
the same cross sections as in Fig.~\ref{sigma}.  The up-sloping dotted
line is the Standard Model muon range.  Horizontal lines, from top
to bottom, correspond to the water equivalent of the Earth along a
diameter, the depth of the ANTARES and AMANDA neutrino observatories,
the thickness of the atmosphere taken horizontally
(along a tangent to the spherical Earth's surface), and the thickness
of the atmosphere taken vertically (zero zenith angle).  The Earth
diameter in cmwe is computed from a parameterized Earth model~\cite{pem}
of Earth's internal structure, and the atmosphere is treated as
exponential, as defined in the text. 
\label{reno}}
\end{figure}

Following a format used previously in the literature~\cite{gqrs},
we can express these cross sections in a more useful manner.
The interaction lengths of neutrinos through various kinds of
matter can be directly compared if they are specified in terms of
the corresponding interaction lengths in a particular medium.
In Fig.~\ref{reno}, the interaction length for neutrinos is
plotted in centimeters of water equivalent (cmwe), indicating the
appropriate energy range necessary for detection underground
and in our atmosphere.  The interaction length in cmwe is 
computed from
\begin{equation}
L_{int}=\frac{1}{\sigma_{\nu N}N_A}\,{\rm ~cmwe},
\label{L_int}
\end{equation}
where $N_A$ is Avogadro's number, since the density of water
is 1 g/cm$^3$.  In the particular case of extra dimension models
with the cross section scaling Eq.~(\ref{sigma_graviton}),
Fig.~\ref{reno} shows that a neutrino becomes unlikely to create a
shower below $5\times10^{18}$~eV, since $M_{4+n}$ can't
be pushed much below about 1 TeV without its effects contradicting
current experiments~\cite{tev-qg}.

The total neutrino-nucleon cross section is dominated
by a contribution of the form Eq.~(\ref{sigma_graviton})
at neutrino energies $E\gtrsim E_{\rm th}$, where, for
$M_{4+n}\gtrsim1$~TeV,
the threshold energy can be approximated by
\begin{equation}
E_{\rm th}\simeq2\times10^{13}\left(\frac{M_{4+n}}
{\rm TeV}\right)^{6.28}\,{\rm ~eV}.
\label{E_thresh}
\end{equation}
Provided that $M_{4+n}$ is low enough, the effects of these
new interactions should be observable in UHE neutrino
interactions. The upper bound Eq.~(\ref{crosslimy}) we derived
on $\sigma_{\nu N}$ by merit of current non-observation of
horizontal air showers immediately implies a corresponding
bound on the mass scale within the context of
Eq.~(\ref{sigma_graviton}):
\begin{equation}
M_{4+n}\gtrsim \bar y^{1/8}
\left(\frac{\phi_c(E)}{10^{-18}\,
{\rm cm^{-2}s^{-1}sr^{-1}}}\right)^{1/4}
\left(\frac{E}{10^{19}\,{\rm eV}}\right)
^{3/8}\,{\rm TeV}\,,\label{Mlimy}
\end{equation}
where $\bar y$ enters because the bound derives
from observational limits at various shower energies.
Since $M_{4+n}$ is a
new constant of nature (that is, the value of $M_{4+n}$
doesn't actually scale with $E$), Eq.~(\ref{Mlimy})
should be evaluated at the energy which generates the
best limit.  The strongest bound that can emerge from
this formula with the lower cosmogenic curve in Fig.~\ref{flux}
occurs at $E=10^{18}$ eV and $\bar y=1$, giving
$M_{4+n}\gtrsim1.2\,{\rm TeV}$.
But for the exchange of Kaluza-Klein gravitons considered
here, we can give a very rough estimate $\bar y\sim0.1$
by noting similarity to massive boson exchange within the
Standard Model~\cite{sigl}. In this case
the best bound is $M_{4+n}(\bar y=0.1)\gtrsim0.9\,{\rm TeV}$.

The bound Eq.~(\ref{Mlimy}) is consistent with the best current
laboratory bound, $M_{4+n}\gtrsim1.26$~TeV, which is from Bhabha
scattering at LEP2~\cite{lep}. The fact that laboratory bounds
are in the TeV range can be understood from Fig.~\ref{sigma}
which shows that for smaller mass scales $\sigma_{tot}$ would
already be dominated by $\sigma_g$ at electroweak scales.
Eq.~(\ref{Mlimy}) can be improved beyond this
level through more data from the continued search for
horizontal showers and independently by any increase in the
theoretically expected UHE neutrino flux beyond the conservative
cosmogenic flux estimate shown in Fig.~\ref{flux}.
We stress again that Eq.~(\ref{Mlimy}) derives from the
more general bound Eq.~(\ref{crosslimy}) valid for
CM energies about 3 orders of magnitude above the electroweak
scale if $\sigma_{\nu N}\propto s$ in this energy range,
and is thus complementary to laboratory bounds.

It is interesting to note that Eq.~(\ref{Mlimy}) does
not depend on the number $n$ of extra dimensions, assuming
the cross section in Eq.~(\ref{sigma_graviton}).  This is in
contrast to some other astrophysical bounds which depend on
the emission of gravitons causing energy loss in stellar
environments~\cite{sn87a,rg}. The best lower limit of this
sort comes from limiting the emission of bulk gravitons into extra
dimensions in the hot core of supernova 1987A~\cite{sn87a}. 
In order to retain the agreement between the energy released by the
supernova, as measured in neutrinos, and theoretical
predictions, the energy flow of gravitons into extra
dimensions must be bounded. The strongest contribution
to graviton emission comes from nucleon-nucleon
bremsstrahlung~\cite{sn87a}. The resulting constraints read
$M_6\gtrsim50\,$TeV, $M_7\gtrsim4\,$TeV, and
$M_8\gtrsim1\,$TeV, for $n=2,3,4$, respectively, and,
therefore, $n\geq4$ is required if neutrino primaries
are to serve as a candidate for the UHECR events observed
above $10^{20}\,$eV (but see~\cite{kp}).  We calculated this
bound for the case $n=7$ extra dimensions, as motivated by
superstring theory.  We find the lower bound drops to
$M_{11}\gtrsim0.05\,$TeV, so for higher numbers of extra
dimensions, the cosmic ray bound derived here could be
stronger than type II supernova bounds, if the scaling
assumed in Eq.~(\ref{sigma_graviton}) were widely accepted.

This assumes that all extra dimensions have the same size given by
\begin{equation}
  r_n\simeq M^{-1}_{4+n}\left(\frac{M_{\rm Pl}}{M_{4+n}}\right)^{2/n}
  \simeq2\times10^{-17}\left(\frac{{\rm TeV}}{M_{4+n}}\right)
  \left(\frac{M_{\rm Pl}}{M_{4+n}}\right)^{2/n}\,{\rm cm}
  \,,\label{rextra}
\end{equation}
where $M_{\rm Pl}$ denotes the Planck mass.  The  SN1987A
bounds on $M_{4+n}$ mentioned above translate into the corresponding
upper bounds $r_6\lesssim3\times10^{-4}\,$mm,
$r_7\lesssim4\times10^{-7}\,$mm, and $r_8\lesssim2\times10^{-8}\,$mm.
With the cosmogenic bound and 7 extra dimensions, we have
$r_{11}\lesssim6\times10^{-12}\,$mm.

A specific signature of the linear scaling with energy of
the cross section Eq.~(\ref{sigma_graviton}) consists of
the existence, for a given zenith angle $\theta$, of two
critical energies.  First, there is an energy $E_1(\theta)$
below which the first interaction point
has a flat distribution in column depth, whereas above
which this distribution will peak above the ground.
This energy is independent of any experimental attributes
of the detector, apart from its altitude.

Second, there is an energy $E_2(\theta)>E_1(\theta)$ below which
a large enough part of the resulting shower lies
within the sensitive volume to be detectable, and
above which the event rate cuts off because primary
neutrinos interact too far away from the detector
to induce observable showers. This energy
is specific to the detector involved.

\begin{figure}
\centering\leavevmode
\epsfig{file=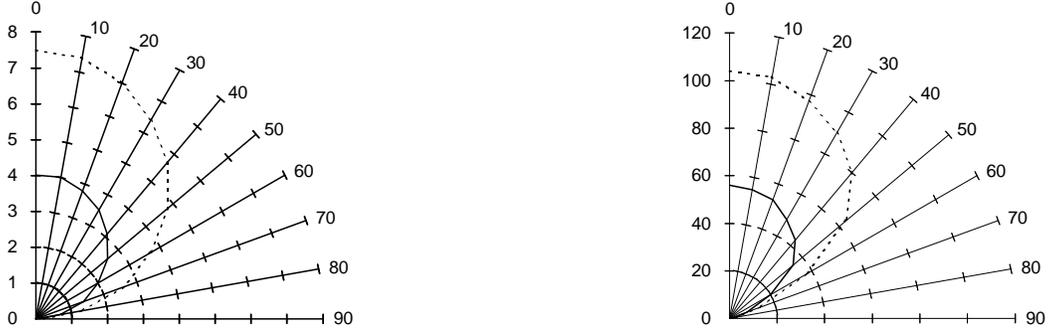,width=6in,clip=,angle=0,
bbllx=193pt,bblly=272pt,bburx=740pt,bbury=488pt}
\vskip1cm
\caption[...]{The critical energies $E_1(\theta)$ (left panel)
and $E_2(\theta)$ (right panel) as a function of zenith angle for
neutrino-induced air showers, as discussed in the text. 
Energy is plotted radially in
units of $10^{20}$~eV.  The solid curve is for $M_{4+n}=1.2$~TeV
and the dashed curve is for 1.4 TeV, for comparison.  These
curves apply to detectors located near an atmospheric depth of
860 g/cm$^2$, including Fly's Eye and both Auger sites.
\label{zenith}}
\end{figure}

We show the zenith angle dependence of these two
critical energies in Fig.~\ref{zenith},
where we have assumed approximate experimental parameters for
the Fly's Eye experiment given in~\cite{baltrusaitis},
at an atmospheric depth of 860~g/cm$^2$; both Auger
sites are within a few hundred meters of this altitude
as well.  For simplicity, we have assumed that the Fly's
Eye detects showers above $10^{19}$~eV, provided that
they initiate and touch ground within 20~km of the detector.  
Following~\cite{bhfpt}, we assume a standard exponential atmosphere of
scale height 7.6 km, and a spherical (rather than planar) Earth.

These zenith angle plots can be expressed in terms of
cross sections instead of energies, and be generalized to other energy
dependences. They have two purposes.  First, upon adding
UHECR data on such a plot, a resemblance to the arcing curves
shown would give strong support to the respective energy scaling.
Second, in the specific case of Eq.~(\ref{sigma_graviton}),
owing to the 4th power dependence on the quantum gravity
scale $M_{4+n}$, Fig.~\ref{zenith}
serves as a good means for determining this mass scale
by comparing the data curves to those predicted. Of course,
this will only work in directions in which the neutrino-induced
EAS rate is not dominated by the ordinary UHECR
induced rate. This restriction may necessitate
additional discriminatory information such as
the shower depth at maximum. Furthermore, in a detailed
analysis the observed shower energy has to be
corrected by the distribution of fractional energy
transfer $y$ in order to obtain the primary neutrino
energy.

This approach requires more data, which will likely come from the
HiRes Fly's Eye and the Pierre Auger observatories.  Because these
facilities can see nitrogen fluorescence shower trails in the sky,
they are better equipped to see horizontal showers than ground
arrays alone.  In addition, they utilize
more than one ``eye'' so their angular resolution is
$\Delta\theta\lesssim2^{\circ}$~\cite{auger-design}.  On the
other hand, these detectors are biased toward horizontal events
because of their long track lengths, and this bias will need to
be treated properly in order to interpret the zenith angle data.

As mentioned above, ``traditional'' neutrino telescopes
based on water and ice as detector medium could be used similarly
to probe a range of smaller $\nu N$ cross sections
$\sim10^{-31}-10^{-29}\,{\rm cm}^2$, if their size significantly
exceeds 1 km and if they are sensitive to at least a limited
range of zenith angles $\theta<90^\circ$, i.e. above the
horizon. For example, the ratio of upgoing to downgoing
events in the $10-100\,$PeV range could be a measure of
the absolute $\nu N$ cross section at these energies~\cite{jain}.
However, the detection method and flavor sensitivity
are different in this case.

It is worth considering whether enhanced cross sections in
extra dimensions significantly raise detector backgrounds.
Here, the most important background for UHE neutrino detection
is from muons created in decays of collision products of
UHECR interactions in the atmosphere (and to a lesser degree,
from tau lepton decays).  The muon range in
centimeters of water equivalent, with only Standard Model
interactions, is approximately~\cite{goobar}
\begin{equation}
L_{\mu}^{SM}\simeq2.5\times10^5\,{\rm ln}\left(
\frac{2E}{{\rm TeV}}+1\right)
\,{\rm cmwe}\,.\label{muon}
\end{equation}
Therefore, the UHE muon range is larger than the depth of the
atmosphere (see Fig.~\ref{reno}).  But with $M_{4+n}\sim$~TeV
extra dimension models, the muon acquires a cross section
given by Eq.~(\ref{sigma_graviton}), and therefore has the same
interaction rate in the atmosphere as the neutrinos we wish
to detect.  However, even if we conservatively estimate the
atmospheric muon flux~\cite{yoshida1}, it is more
than 350 times smaller than the cosmogenic neutrino flux at
$E=10^{18}$~eV, and more than $10^3$ times smaller for
$E\geq10^{19}$~eV.  So with extra dimensions, atmospheric muons
may lead to interesting signatures in atmospheric and underground
detectors, but they do not hinder cosmic UHE neutrino shower
identification.

UHECRs and neutrinos together with other astrophysical
and cosmological constraints thus provide an interesting testing
ground for new interactions beyond the Standard Model, as
suggested for example in scenarios involving
additional large compact dimensions. In the context of these
scenarios we mention that Newton's law of gravity is expected to be
modified at distances smaller than the length scale given by
Eq.~(\ref{rextra}). Indeed, there are laboratory 
experiments measuring the gravitational interaction at smaller
distances than currently measured (for a recent review of
such experiments see Ref.~\cite{lcp}), which also probe
these theories. Thus, future UHECR experiments and
gravitational experiments in the laboratory together 
have the potential to provide rather strong tests of these theories. 
These tests complement constraints
from collider experiments~\cite{coll-extra-dim}.

\section{Conclusions}

A direct measurement approach has been presented here for bounding
the extent to which UHE neutrinos acquire greater interactions with
matter than the Standard Model prescribes.  By looking for 
nearly horizontal air showers, we can limit the UHE neutrino flux
striking Earth.  When this is combined with a natural
source of cosmogenic UHE neutrinos, a model-independent upper
limit on the cross section results. Air shower experiments can
also test for the energy dependence of cross sections in the
range $\sim10^{-29}-10^{-27}\,{\rm cm}^2$; we have discussed the
specific case of a linear energy dependence, as motivated
by some large extra compact dimensions models that have enjoyed
much recent attention in the community. In such scenarios,
fundamental quantum gravity scales in the $1-10\,$TeV range
can be probed.  This could also be relevant for testing string
theory scenarios with string scales in the TeV range, however
reliable theoretical predictions for the relevant cross sections
in such scenarios are not yet available.

If this method proves beneficial, knowledge will be gained
about new interactions at ultra-high energies.
Should the data support extra dimensions, our understanding of
UHE interactions between {\it all} particles (not just $\nu N$)
will be changed because changes in gravity affect
every particle. In general, cosmic neutrinos and UHECRs
will probe interactions at energies a few orders of magnitude
beyond what can be achieved in terrestrial accelerators in
the forseeable future.

\section*{Acknowledgments}

We would like to thank Pierre Binetruy, Sean Carroll, Cedric
Deffayet, Emilian Dudas, Gia Dvali, and Michael Kachelriess
for valuable conversations, and Mary Hall Reno and Ren-Jie Zhang
for their timely correspondence.  We also thank Murat Boratav,
Shigeru Yoshida, and Tom Weiler for comments on the manuscript.

\vfill\eject

\vfill\eject

\vfill\eject

\vfill\eject
%\vskip1cm

\end{document}